\documentclass[showpacs,aps,prb,twocolumn,superscriptaddress]{revtex4-1}
\usepackage{graphicx} 
\usepackage{dcolumn}
\usepackage{bm}
\usepackage{amssymb,amsmath}
\usepackage{epstopdf}

\begin{document}
\title{Electromagnetically induced transparency control in terahertz metasurfaces based on bright-bright mode coupling}

\normalsize
\author{R.~Yahiaoui}
\email[]{riad.yahiaoui@howard.edu}
\thanks{corresponding author.}
\affiliation{Department of Physics and Astronomy, Howard University, Washington, DC 20059, USA}
\author{J. A.~Burrow}
\affiliation{Electro-Optics Department, University of Dayton, Dayton, OH 45469, USA}
\author{S. M.~Mekonen}
\affiliation{Department of Physics and Astronomy, Howard University, Washington, DC 20059, USA}
\author{A.~Sarangan}
\affiliation{Electro-Optics Department, University of Dayton, Dayton, OH 45469, USA}
\author{J.~Mathews}
\affiliation{Department of Physics, University of Dayton, Dayton, OH 45469, USA}
\author{I.~Agha}
\affiliation{Electro-Optics Department, University of Dayton, Dayton, OH 45469, USA}
\affiliation{Department of Physics, University of Dayton, Dayton, OH 45469, USA}
\author{T. A.~Searles}
\email[]{thomas.searles@howard.edu}
\thanks{corresponding author.}
\affiliation{Department of Physics and Astronomy, Howard University, Washington, DC 20059, USA}
\date{\today}

\begin{abstract}
We demonstrate a classical analogue of electromagnetically induced transparency (EIT) in a highly flexible planar terahertz metamaterial (MM) comprised of three-gap split ring resonators. The keys to achieve EIT in this system are the frequency detuning and hybridization processes between two bright modes coexisting in the same unit cell as opposed to bright-dark modes. We present experimental verification of two-bright mode coupling for a terahertz EIT-MM in the context of numerical results and theoretical analysis based on a coupled Lorentz oscillator model. In addition, a hybrid variation of the EIT-MM is proposed and implemented numerically in order to dynamically tune the EIT window by incorporating photosensitive silicon pads in the split gap region of the resonators.  As a result, this hybrid MM enables the potential active optical control of a transition from the on-state (EIT mode) to the off-state (dipole mode).
\end{abstract}


\maketitle


Metamaterials (MMs) offer versatile and remarkable ways to manipulate electromagnetic waves for extraordianary applications such as sub-diffraction focusing~\cite{fang2005sub}, sensing~\cite{riad2016thz,cong2015experimental}, electro-magnetic cloaking\cite{schurig2006meta}, near-perfect absorption~\cite{riad2015trap,riad2017broad}, etc. In recent years, the concept of electromagnetically induced transparency (EIT) has attracted significant interest due to its potential applications in sensing, controllable delay lines, optical buffers, slow light devices and nonlinear effects~\cite{liu2009planar,longdell2005stop,totsuka2007slow,yanik2004stop}. EIT is a quantum phenomenon that arises from the destructive interference between different excitation pathways in a three level atomic system, making an initially opaque medium transparent to a probe laser beam~\cite{harris1990nl,fleischhauer2005elec}.  However, practical applications of EIT are limited by severe experimental conditions such as high intensity lasers and cryogenic temperatures.  The emergence of MMs as classical analogues of EIT have paved the way for the development of new applications such as sensing, THz modulators and slow light metadevices for operation at room temperature.

The generation of an EIT analogue in MMs is achieved by two different approaches: (i) bright-dark mode coupling and (ii) bright-bright mode coupling. The first approach typically involves a bright mode resonator coupled to the incident wave directly which is highly radiative and exhibits a low Q-factor.  By contrast, the dark mode resonator is characterized by a high Q-factor, and could not be excited by the incident wave directly. However, this dark mode is activated by the bright mode resonator via near-field coupling. In such systems, the necessary condition to achieve EIT is to couple the bright and quasi-dark resonances at the same resonance frequency with each possessing contrasting linewidths~\cite{singh2009coupling,rao2017modulating,manjappa2017mag,devi2017pit,zhang2008pit,riad2017active,xu2016freq,gu2012active}. 

The second approach based on the frequency detuning and hybridization of two bright modes placed in a close proximity of one another has been rarely reported in literature. Such reported works for bright-bright mode coupling in the terahertz frequency regime is very limited~\cite{chiam2009bright,qiao2015bright,li2011bright,ma2011bright}. In one example, the authors have demonstrated a plasmon induced transparency (PIT) by hybridizing two concentric-twisted double split ring resonators (DSRRs) under various geometrical configurations on flexible polyimide substrates~\cite{hokmabadi2015pit}. And recently, a graphene ring/strip composite planar design is proposed to achieve numerically a THz PIT effect through bright-bright mode coupling~\cite{zhang2017novel}. 

Here, we present a detailed study of the hybridization process between two bright modes of a single unit cell and show active control of the EIT effect in the terahertz. For this case, we removed the bottom capacitive gap of a polarization independent four-gap square split ring resonator~\cite{burrow2018bio}.  This paper is organized as follows.   We first propose the design of the unit cell and give experimental details for a novel and simple planar MM structure to demonstrate the EIT effect in the terahertz (THz) frequency regime.  Then, we introduce experimental results supported by numerical solutions in the context of a two-coupled oscillator analytical model.  To demonstrate active control within this system, we present the concept of an optically reconfigurable EIT effect associated with large group delays.  

The unit cell of the proposed EIT-MM is schematically shown in the inset of Fig. 1 (top left). It is composed of an array of metallic three-gap split ring resonators (TGSRRs), deposited periodically on the top side of 50.8 $\mu m$ thick and highly flexible polyimide substrate.   Recently, there have been various efforts devoted to demonstrate flexible metamaterials~\cite{burrow2018bio,burrow2017pol,riad2015multi,riad2013ultra,riad2014thz}. The use of flexible substrates has provided an unprecedented route to achieve active tunability in the frequency of MMs due to modifications in the profiles and the periodicities of the structures when the substrates are stretched~\cite{chen2015flex,tao2008thz,lee2012reverse,li2013mech,zhang2015mech}.  

The TGSRRs were made from 100 nm thick Ag patterned using standard photolithography methods resulting in a planar metasurface as shown in Fig. 1. The active surface area of the fabricated device is about 1 cm  x 1 cm and the relevant geometrical dimensions of the unit cell are: $p_{x} = p_{y} =  300~\mu$m, $l =  250~\mu$m, $w =  35~\mu$m, and $g_{1} = g_{2} = g_{3} = g = 35~\mu$m.  Such periodic structures do not diffract normally incident electromagnetic radiation for frequencies less than 1 THz. 
\begin{figure}
\includegraphics [scale = 0.35] {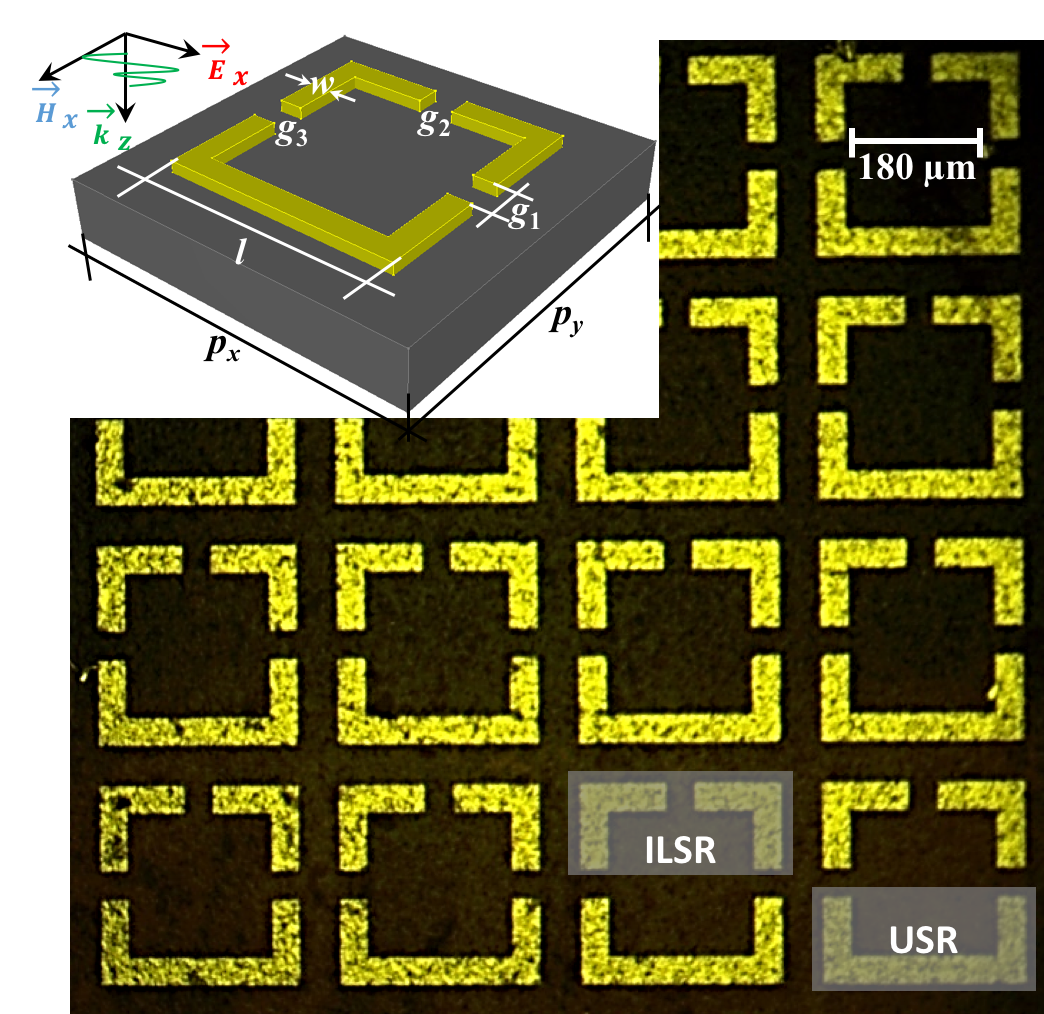}
\caption{Optical micrograph of the fabricated TGSRRs metasurface composed of inverted L-shape resonators (ILSRs) and U-shaped resonators (USRs). The unit cell is illustrated schematically in the inset (top left) with the corresponding electromagnetic excitation configuration. The relevant geometrical dimensions are: $p_{x} = p_{y} =  300~\mu$m, $l =  250~\mu$m, $w =  35~\mu$m, and $g_{1} = g_{2} = g_{3} = g = 35~\mu$m.}
\label{Figure1}
\end{figure}

Numerical calculations were carried out using a finite element method (FEM). 
In these calculations, the elementary cell of the designed metasurface was irradiated at normal incidence, under TM-polarization ($E\parallel  x$-axis), as indicated in the inset of Fig. 1. Periodic boundary conditions were applied in the numerical model in order to mimic a 2D infinite structure. In simulations, the polyimide film was treated as a dielectric with $\varepsilon_{sub} = 3.3 + i0.05$ and the silver (Ag) was modeled as a lossy metal with a conductivity of $6.1\times10^{7}$ S/m.

Measurements of the TGSRRs were performed using linearly polarized collimated radiation from a continuous-wave (CW) THz spectrometer (Teraview CW Spectra 400). The nominal spectral resolution of 100 MHz is governed by the precision of the temperature tuning technique of two near-IR diode lasers and not by a mechanical delay stage as found in a conventional THz time-domain spectroscopy (TDS) setup.   The transmission spectrum from the sample was determined as $T(f) = P_{MM}(f)/P_{sub}(f)$, where $P_{MM}(f)$ and $P_{sub}(f)$ are the filtered THz power spectra of the planar metamaterial and flexible substrate respectively.

\begin{figure}[h]
\includegraphics [scale = 0.64] {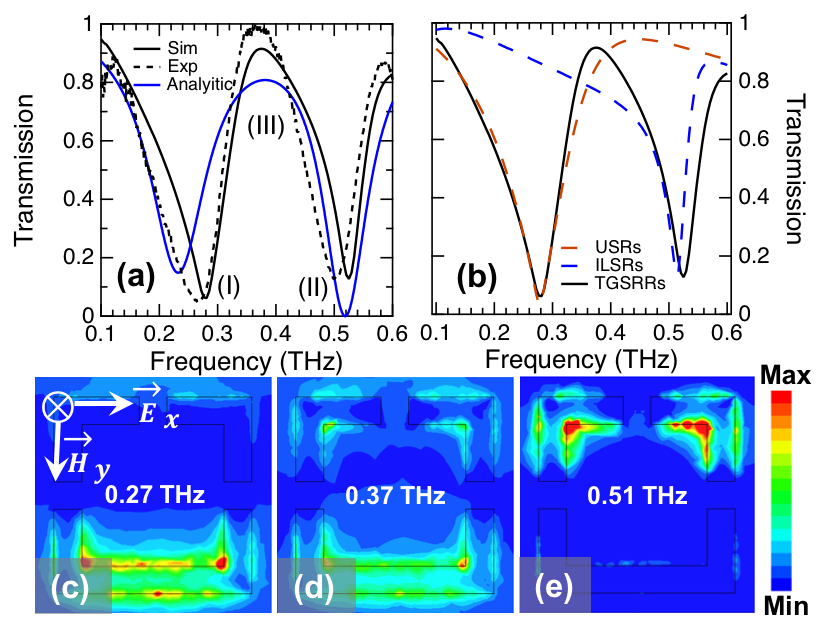}
\caption{(a) Simulated (solid black line) and measured (dashed black line) terahertz transmission spectrum of the EIT-MM. The blue solid curve represents the analytically fitted data (T = 1 - Im($\chi$)) using the two-oscillator model (Eqn. 6). (b) Comparison of transmission spectra between TGSRRs EIT structure and its constituent elements USR and ILSRs, respectively. (c)-(e) Spatial distribution of the resonant magnetic field $H_{z}$ in a single unit cell consisting of two radiative subresonators, calculated at 0.27 THz, 0.37 THz and 0.51 THz, respectively.}
\label{Figure2}
\end{figure}

The simulated (solid black line) and measured (dashed black line) transmission spectra of the TGSRRs metasurface are plotted in Fig. 2(a) and are in good agreement with each other. A typical EIT-like spectral signature is observed, where a transmission peak appears at about 0.37 THz with an amplitude close to unity between two resonance dips at around 0.27 THz and 0.51 THz, respectively. Under the illumination of a linearly polarized THz wave ($E_{x}$), each unit cell of the EIT-MM can be regarded as the combination of a lower U-shaped resonator (USR) and an upper pair of inverted L-shape resonators (ILSRs) placed face-to-face (see Fig. 1, bottom right).

The EIT effect can be interpreted by analyzing numerically the resonances of each constituent TGSRR as well as their interactions.  Figure~2(b) shows the transmission spectrum of the EIT-MM (solid black line) superimposed with that of its two constituent components USR (dashed red line) and ILSRs (dashed blue line), respectively. One can clearly observe that the two transmission dips stem from the resonances associated with each individual subresonator (USR and ILSRs), and a transparency window appears at 0.37 THz due to interaction between the USR and ILSRs after hybridization. 

This behavior is further analyzed by simulating the z-components of the magnetic field distributions, as shown in Figs. 2(c)-(e); at two transmission dips I (0.27 THz) and II (0.51 THz) and a transmission peak III (0.37 THz). The USR is strongly excited by the incident wave at resonance I (0.27 THz), while the magnetic field primarily concentrates on the ILSRs at resonance II (0.51 THz).  At the transparency frequency III (0.37 THz), we can see both elements are excited due to the frequency detuning at 0.37 THz.  Furthermore, this resonance detuning makes the magnetic field distribution of the transmission peak smaller than that of the transmission dips, as shown in Figs. 2(c)-2(e). Therefore, it becomes evident that the transparency window of the EIT-MM is a direct consequence of spectral combinations of USRs and ILSRs verifying our findings in Fig. 2(b). Interestingly, unlike conventional EIT phenomena where bright and dark modes possess contrasting linewidths resonating at the same resonance frequency, the EIT demonstrated here is generated by coupling two bright modes with comparable bandwidths. 

To demonstrate the validity of the underlying EIT effect, we used an analytical model based on the coupled oscillator theory described by the following set of equations~\cite{riad2017active,meng2012pol}: 
\begin{equation}
\ddot{x}_{a}(t) + \gamma_{a}\dot{x}_{a}(t) + \omega_{a}^{2}{x}_{a}(t)+\Omega^{2}{x}_{b}(t) =\frac{Q}{M}E
\end{equation}
\begin{equation}
\ddot{x}_{b}(t) + \gamma_{b}\dot{x}_{b}(t) + \omega_{b}^{2}{x}_{b}(t)+\Omega^{2}{x}_{a}(t) =\frac{q}{m}E.
\end{equation}
Here, USR and ILSRs are designated as particles $a$ and $b$, respectively $(Q,q)$, $(M,m)$, $(\omega_{a},\omega_{b})$ and $(\gamma_{a},\gamma_{b})$  are the effective charges, effective masses, resonance angular frequencies and the loss factors of the particles.  $\Omega$ defines the coupling strength between the particles. Here, we consider both particles interact with the incident THz electric field $E=E_{0}e^{i\omega t}$. In the above coupled equations, we substitute $q=Q/A$ and $m=M/B$, where $A$ and $B$ are dimensionless constants that dictate the relative coupling of incoming radiation with the particles. Now by expressing the displacements vectors for particles $a$ and $b$ as $x_{a}=c_{a}e^{i\omega t}$ and $x_{b}=c_{b}e^{i\omega t}$, we solve the above coupled equations (1) and (2) for $x_{a}$ and $x_{b}$:
\begin{equation}
{x}_{a}=\frac{\frac{B}{A}\Omega^{2}+(\omega^{2}-\omega_{b}^{2}+i\omega \gamma_{b})}{\Omega^{4}-(\omega^{2}-\omega_{a}^{2}+i\omega \gamma_{a})(\omega^{2}-\omega_{b}^{2}+i\omega \gamma_{b})}\frac{Q}{M}E
\end{equation}
\begin{equation}
{x}_{b}=\frac{\Omega^{2}+\frac{B}{A}(\omega^{2}-\omega_{a}^{2}+i\omega \gamma_{a})}{\Omega^{4}-(\omega^{2}-\omega_{a}^{2}+i\omega \gamma_{a})(\omega^{2}-\omega_{b}^{2}+i\omega \gamma_{b})}\frac{Q}{M}E.
\end{equation}
The susceptibility $\chi$, which relates the polarization ($P$) of the particle to the strength of incoming electric field ($E$) is then expressed in terms of the displacement vectors as: 
\begin{equation}
\chi=\frac{P}{\varepsilon_{0}E}=\frac{Q x_{a}+ q x_{b}}{\varepsilon_{0}E}
\end{equation}
\begin{equation}
\begin{aligned}
\chi={}&\frac{K}{A^{2}B}\bigg(\frac{A(B+1)\Omega^{2}+A^{2}((\omega^{2}-\omega_{b}^{2})+B(\omega^{2}-\omega_{a}^{2}))}{\Omega^{4}-(\omega^{2}-\omega_{a}^{2}+i\omega \gamma_{a})(\omega^{2}-\omega_{b}^{2}+i\omega \gamma_{b}}\\
&+i\omega \frac{A^{2}\gamma_{a}+B\gamma_{b}}{\Omega^{4}-(\omega^{2}-\omega_{a}^{2}+i\omega \gamma_{a})(\omega^{2}-\omega_{b}^{2}+i\omega \gamma_{b}}\bigg).\\
\end{aligned}
\end{equation}    

The experimental transmission in Fig. 2(a) is fitted by the imaginary part of the nonlinear susceptibility expression.  In our fitting, the transmission coefficient is defined as T = 1-Im($\chi$) (given by the Kramer-Kronig relations), which is derived from the conservation of energy relation T + A = 1 (normalized to unity), where A = Im($\chi$) is the absorption (losses) within the medium.  For the fit, the values of the loss factors and are obtained from the linewidths of the curves shown in Fig. 2(a), which are calculated to be around $7.35 \times 10^{11}$ rad/s and $5.71 \times 10^{11}$ rad/s, respectively. Then, we used $\omega_{a}=\sqrt{\omega_{o}^{2}-\Omega^{2}}$  to get  $1.58 \times 10^{12}$ rad/s. By substituting the calculated values for $\gamma_{a}$, $\gamma_{b}$, $\omega_{a}$, $\omega_{b}$, $\Omega$ and by setting A=1.5 and B=1.25 (mass of USR is about 1.25 times the mass of ILSRs), we plotted the analytically modeled transmission coefficient in Fig. 2(a), which exhibits good agreement with the corresponding experimentally measured and numerically simulated curves.  Although there are minor differences in amplitude and bandwidths for the spectra in Fig. 2(a), the spectrum obtained from the numerical model validates the experimental data.  Slight differences can be associated with small imperfections in the fabrication process, dispersions being unaccounted for during simulations, etc.  

In Fig. 3, we actively change the conductivity of the dielectric gap $g_{2}$ for the resonators to dynamically modulate the amplitude of the EIT window and switch between different operation regimes, from the on-state (EIT mode) to the off-state (dipole mode). To this end, 100 nm-thick photosensitive silicon pads are integrated into the split gap  $g_{2}$ of the resonators. The hybrid TGSRRs/silicon EIT-MM [see Fig. 3(a)] forms a tunable resonating structure that can, upon application of external optical power, actively modify the resonance strength and resonance mode.

\begin{figure}
\includegraphics [scale = 0.75] {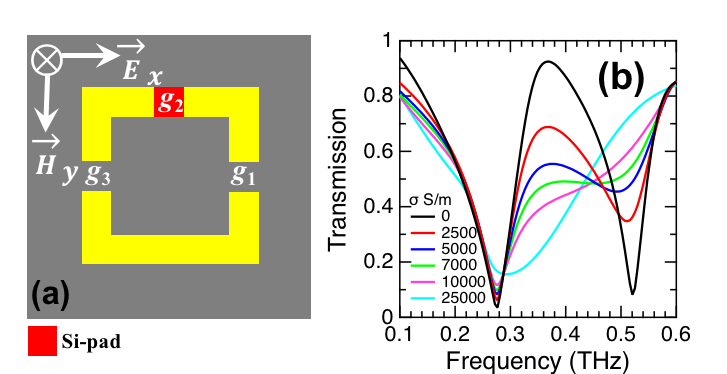}
\caption{(a) Schematic view of the optically controlled EIT-MM unit cell. The yellow regions are Ag, the grey region is polyimide film and the red region is photosensitive silicon. (b) Evolution of the simulated transmission spectra of the tunable EIT metamaterial with varying the conductivity $\sigma_{Si}$ of the photoconductive silicon pads in the range  $0 - 2.5 \times 10^{4}$ S/m.}
\label{Figure3}
\end{figure}
Numerically, we applied a simple conductivity model for the silicon, considering a constant permittivity of Si = 11.7 and a pump-power-dependent photoconductivity Si varying from 0 S/m to $2.5 \times 10^{4}$ S/m. As the conductivity of the Si pad is gradually increased up to $2.5 \times 10^{4}$ S/m, the transparency window undergoes a gradual reduction in its amplitude and finally degenerates to a single broad dipole-like resonance.  This resonance completely annihilates the EIT effect in the system leading to a modulation depth (MD) of $\sim$77\% (Fig. 3(b)).  Note that the MD depends on the sensitivity of the resonance and the photoconductive response of the deposited silicon patch. Therefore, active modulation of the EIT window is caused by the photoexcitation of free carriers in the silicon pads integrated into the capacitive gaps, hence gradually weakening the strength of the ILSRs resonance in the system.
\begin{figure}
\includegraphics [scale = 0.36] {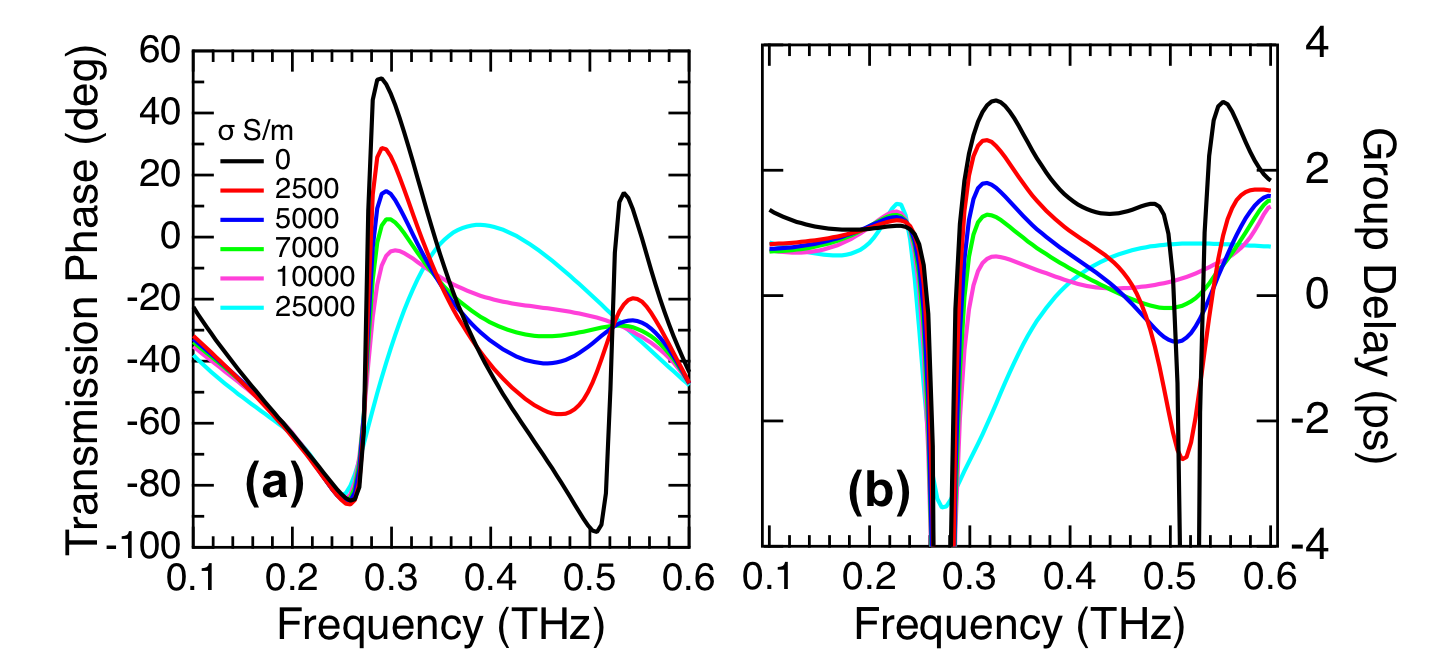}
\caption{(a) The simulated transmission phase and (b) group delay of the EIT metamaterial with increasing the conductivity $\sigma_{Si}$ of the photoconductive silicon pads in the range $0 - 2.5 \times 10^{4}$ S/m.}
\label{Figure1}
\end{figure}

A remarkable characteristic of EIT response is the ability to achieve strong dispersion and slow light effects. A controllable slow-light metamaterial can trap photons for a long time inside the structure which is useful to enhance light-matter interactions. The measurement of the slow light effect is represented by the group delay ($\tau_{g}$) of the incident THz wave packet through the sample, calculated using $\tau_{g}$ = -d$\phi(\omega)$/d$\omega$, $\omega = 2\pi$f, where $\phi(\omega)$ and f are transmission phase and frequency, respectively. For the hybrid EIT-MM presented in Fig. 3(a), one can observe a strong phase dispersion around the transmission window causing a large $\tau_{g}$, as shown in Figs. 4(a) and 4(b). 	The group delay retrieved from the numerical data of TGSRRs is plotted in Fig. 4(b).  One can observe that at the resonance frequencies of 0.27 THz and 0.51 THz, the metamaterial exhibits negative group delay. In the vicinity of the transparency peak, large positive group delays are obtained, indicating potential use in slow light applications. For example, at around 0.33 THz where more than 80\% of EIT transmission is achieved, the THz radiation experiences a delay of about 3.18 ps, corresponding to the time delay of a 954 $\mu$m distance of free space propagation [Fig. 4(b), black curve]. 

The dependence of the transmission phase spectra and group delays with various silicon conductivities are also depicted in Figs. 4(a) and 4(b), respectively. As the conductivity of the silicon pads increases to $2.5 \times 10^{4}$ S/m, the EIT metamaterial array gradually loses its slow light characteristic; finally turning into a typical dipole-like group delay feature. Therefore, we can achieve the capability of switching the group delay and controlling the amount of the group delay by tuning the conductivity of the silicon.  This capability can be strategically important in designing very compact slow light devices with ultrafast response.

In summary, we have designed, fabricated and experimentally characterized a classical analogue of EIT for a highly flexible MM in the THz regime. The observed EIT effect was modeled using coupled oscillators, which showed good agreement with the observed results. Moreover, the integration of photosensitive silicon pads in the numerical model leads to an active control of the EIT window. The simple topology and low cost of our proposed structures are key parameters suggest a promising route towards mass production of a wide array of applications including THz modulators, biosensing and slow-light devices. 

\begin{acknowledgments}
Funding for this research comes from the Air Force Office of Scientific Research (FA9550-16-1-0346) and the NSF (No. 1541959, No. 1710273, No. 1709200).  T. A. S. acknowledges support from the CNS Scholars Program, S. M. M. acknowledges support in the form of the Just-Julian Graduate Research Assistantship and J. A. B. would like to thank AFRL Minority Leadership program for support.
\end{acknowledgments}


\end{document}